# Hamilton-Jacobi theory for Hamiltonian and non-Hamiltonian systems

Sergey A. Rashkovskiy

*Institute for Problems in Mechanics of the Russian Academy of Sciences, Vernadskogo Ave., 101/1, Moscow, 119526, Russia*

*Tomsk State University, 36 Lenina Avenue, Tomsk, 634050, Russia*

E-mail: rash@ipmnet.ru, Tel. +7 906 0318854

**Abstract**

A generalization of the Hamilton-Jacobi theory to arbitrary dynamical systems, including non-Hamiltonian ones, is considered. The generalized Hamilton-Jacobi theory is constructed as a theory of ensemble of identical systems moving in the configuration space and described by the continual equation of motion and the continuity equation. For Hamiltonian systems, the usual Hamilton-Jacobi equations naturally follow from this theory. The proposed formulation of the Hamilton-Jacobi theory, as the theory of ensemble, allows interpreting in a natural way the transition from quantum mechanics in the Schrödinger's form to classical mechanics.

## 1. Introduction

In modern physics, the Hamilton-Jacobi theory occupies a special place. On the one hand, it is one of the formulations of classical mechanics [1-3], while on the other hand it is the Hamilton-Jacobi theory that is the classical limit into which quantum (wave) mechanics passes at $\hbar \to 0$. In particular, the Schrödinger equation describing the motion of quantum particles, in the limit $\hbar \to 0$, splits into two real equations: the classical Hamilton-Jacobi equation and the continuity equation for some ensemble of classical particles [4]. Thus, it is the Hamilton-Jacobi theory that is the bridge that connects the classical mechanics of point particles with the wave mechanics of quantum objects.

From the standpoint of classical physics, the Newtonian, Lagrangian, or Hamiltonian formulations of classical mechanics that describe the motion of individual point particles are more "natural" than the Hamilton-Jacobi theory which uses the continual description. In this regard, it remains unclear why does quantum mechanics, in the limit $\hbar \to 0$, turn into classical mechanics precisely in the Hamilton-Jacobi form, but not in its other possible forms?

Note that numerous attempts to build the quantum mechanics of an individual particle using the classical concepts such as "velocity", "classical trajectory" etc. were unsuccessful [5,6]. This also applies to the so-called Bohmian mechanics [7-9], which cannot be considered a completely classical formulation of quantum mechanics, since the velocity field in which the motion of the individual particles is calculated is found by solving the Schrödinger equation. For this reason, it is more correct to call the Bohmian mechanics by the Bohmian kinematics, while the corresponding dynamics is described by the Schrödinger equation.

It can be hoped that a deeper understanding the meaning of classical Hamilton-Jacobi theory will allow a deeper understanding the quantum mechanics, which, according to the Copenhagen interpretation, is in fact a field theory describing the "fields of probability".

The Hamilton-Jacobi theory is traditionally understood in the narrow sense [1-3] as one of the formulations of classical mechanics, in which not an individual, but a continual description of the motion of particles is used. Unlike the "individual" formulations of classical mechanics that describe the motion of a particle using a system of ordinary differential equations, the Hamilton-Jacobi theory contains a single scalar partial differential equation of the first order - the Hamilton-Jacobi equation. Herewith, the trajectories of individual particles are the characteristics of this equation, while the transition to the motion of individual particle, in the Hamilton-Jacobi theory, occurs using the Jacobi theorem [1, 2].

There are different ways of the derivation of the Hamilton-Jacobi equation. The first derivation was proposed by Hamilton himself, who applied the methods of geometric optics to the solution of mechanical problems. As a result, he derived his equation, based on the similarity of the laws of mechanics with the laws of geometric optics, the study of which prompted Hamilton to develop his theory. At present, Hamilton's approach, in fact, underlies the physical justification for the transition from quantum mechanics to classical mechanics: the Hamilton-Jacobi equation is the eikonal equation for the Schrödinger equation, which goes over into it in the limit of small wavelengths.

Another more formal derivation of the Hamilton-Jacobi equation, expounded in all textbooks on mechanics, is based on the definition of the action or on the properties of canonical transformations [1-3]: the action or generating functions in this approach satisfy the Hamilton-Jacobi equation. Such an approach formalizes the derivation of the Hamilton-Jacobi equation, but makes it difficult to understand the transition from quantum mechanics to classical mechanics. Moreover, such an approach is applicable only to Hamiltonian systems, and cannot be generalized to non-Hamiltonian systems.

We note that speaking of the Hamilton-Jacobi theory, one often has in mind the methods for solving the problems on the motion of classical systems based on the solution of the Hamilton-Jacobi equation.

Under the influence of quantum mechanics, in some papers, the Hamilton-Jacobi theory began to be interpreted in broader sense including in it not only the Hamilton-Jacobi equation itself, but also the continuity equation describing the distribution of an ensemble of identical systems in the configuration space [5-15]. In other words, the Hamilton-Jacobi theory is implicitly considered as a theory of an ensemble of identical mechanical systems that differ only in the initial conditions. As far as we know, such a view of the Hamilton-Jacobi theory has not been specifically investigated by anyone. At the same time, it seems to be important both from the point of view of understanding the classical Hamilton-Jacobi theory itself and from the point of view of understanding the connection between quantum and classical mechanics.

In this paper we will understand the Hamilton-Jacobi theory in an extended sense as the theory of an ensemble. In such an approach, the motion of the ensemble is described in natural way by the continual equation of motion and the continuity equation in the configuration space, while the Hamilton-Jacobi theory in the generalized sense becomes a natural formulation of classical mechanics, along with the Newtonian, Lagrangian, Hamiltonian, etc. Such a view on the Hamilton-Jacobi theory allows generalizing it in natural way to arbitrary including non-Hamiltonian systems.

We show that the Hamilton-Jacobi theory in the traditional (narrow) sense is a special case of the theory considered here. We give a number of examples showing that in the case of classical Hamiltonian systems, the proposed theory leads to the well-known Hamilton-Jacobi equations. Moreover, the proposed theory provides a general method for deriving the Hamilton-Jacobi equations for any Hamiltonian systems, regardless of their physical nature.

## 2. Hamilton-Jacobi ensemble

Let us introduce the Hamilton-Jacobi ensemble [15] using the example of a system consisting of a single point particle of mass $m$ moving in space under the action of an external force $\mathbf{F}(t,\mathbf{r},\mathbf{v})$.

The motion of such a particle is described by the Newton's equation

$$m\frac{d\mathbf{v}}{dt} = \mathbf{F}(t,\mathbf{r},\mathbf{v}) \qquad (1)$$

which is solved together with the kinematic equation

$$\frac{d\mathbf{r}}{dt} = \mathbf{v} \qquad (2)$$

which is the definition of the velocity of the particle.

In the general case, such a system can be non-Hamiltonian.

The solution of the system of equations (1), (2) depends on the initial conditions: particles with different initial conditions move in space along the different trajectories.

Let us consider the set of identical non-interacting classical particles that differ only in the initial conditions. Although all the particles under consideration move in the same physical space, when describing an ensemble we will talk about the configuration space $\{q\}$ which is a conditional space in which all the particles of the ensemble move simultaneously, but they do not interact each other. Let us impose on the ensemble an additional condition of single-valuedness: all particles that are at the same point in the configuration space at the same time must have the same velocity.

Such an ensemble we will calle the Hamilton-Jacobi ensemble.

The Hamilton-Jacobi ensemble can be regarded as a compressible continuous medium moving in an external force field $\mathbf{F}(t, \mathbf{r}, \mathbf{v})$. By analogy with hydrodynamics, two approaches to describing the motion of the Hamilton-Jacobi ensemble are possible: Lagrangian and Eulerian [16]. The Lagrangian approach considers the motion of each (Lagrangian) particle of a continuous medium, which for the Hamilton-Jacobi ensemble coincide with physical particles, and observes their motion in space. In the Eulerian approach, the Hamilton-Jacobi ensemble is considered as a continuous medium, which is described by the velocity field $\mathbf{v}(t, \mathbf{r})$ and density $\rho(t, \mathbf{r})$, herewith the velocity of the Hamilton-Jacobi ensemble at the point $\mathbf{r}$ coincides with the velocity of those particles that are at this point at a given time.

Similar to hydrodynamics [16], for the Hamilton-Jacobi ensemble, one can introduce the concepts of the material (Lagrangian) time derivative $\frac{d}{dt}$ describing the time variations associated with an individual particle and the local time derivative $\frac{\partial}{\partial t}$ describing the time variation in a fixed point of space.

Material and local derivatives with respect to time are related by the expression [16]

$$\frac{d}{dt} = \frac{\partial}{\partial t} + (\mathbf{v}\nabla) \qquad (3)$$

In the Lagrange approach, the equations of motion are equations (1) and (2). In the Eulerian approach, the equation of motion is obtained by substituting the material derivative (3) into equation (1). As a result, we obtain the Euler equation

$$m\left(\frac{\partial \mathbf{v}}{\partial t}+(\mathbf{v}\nabla)\mathbf{v}\right)=\mathbf{F}(t,\mathbf{r},\mathbf{v}) \tag{4}$$

For the Hamilton-Jacobi ensemble, as for any continuous medium in the Eulerian approach, the continuity equation can be introduced

$$\frac{\partial \rho}{\partial t}+\operatorname{div}\rho\mathbf{v}=0 \tag{5}$$

The density $\rho$ can be normalized in various ways, for example, it can be considered as the mass or numerical (the number of particles per unit volume of the configuration space) density of the Hamilton-Jacobi ensemble; it can also be normalized to unity and considered, as is done in quantum mechanics, as a probability density of finding the particle at a given point in space.

The system of equations (4), (5), describing the motion of the Hamilton-Jacobi ensemble in the configuration space, will be called the Hamilton-Jacobi theory in the generalized sense.

We claim that the equation of motion (4) is an analog of the Hamilton-Jacobi equation for arbitrary (both Hamiltonian and non-Hamiltonian) systems, and equation (4), together with the continuity equation (5) generalize the Hamilton-Jacobi theory (which is understood here as the theory of the Hamilton-Jacobi ensemble) to arbitrary, including non-Hamiltonian, systems. In order to demonstrate this, we will consider two examples: one for the Hamiltonian system, and the other for the non-Hamiltonian system.

### 3. Nonrelativistic motion of a particle in a potential field

Let us consider the motion of a particle in external potential field $U(t,\mathbf{r})$:

$$\mathbf{F}(t,\mathbf{r})=-\nabla U \tag{6}$$

In this case, taking into account the identity

$$(\mathbf{v}\nabla)\mathbf{v}=\frac{1}{2}\nabla\mathbf{v}^2-[\mathbf{v}\operatorname{rot}\mathbf{v}] \tag{7}$$

one writes the equation (4) in the form

$$m\left(\frac{\partial \mathbf{v}}{\partial t}-[\mathbf{v}\operatorname{rot}\mathbf{v}]\right)=-\nabla\left(\frac{1}{2}m\mathbf{v}^2+U\right) \tag{8}$$

Equation (8) has a solution in the form of a potential flow of the Hamilton-Jacobi ensemble

$$\mathbf{v}(t,\mathbf{r})=\frac{1}{m}\nabla S(t,\mathbf{r}) \tag{9}$$

where the function $\frac{1}{m}S(t,\mathbf{r})$ plays a role of the velocity potential.

In this case the equation (8) takes the form

$$\nabla\left(\frac{\partial S}{\partial t} + \frac{1}{2}m\mathbf{v}^2 + U\right) = 0 \tag{10}$$

Then, taking (8) into account, one obtains the Hamilton-Jacobi equation for the function $S(t,\mathbf{r})$:

$$\frac{\partial S}{\partial t} + \frac{1}{2m}|\nabla S|^2 + U(t,\mathbf{r}) = 0 \tag{11}$$

We note that, from the point of view of hydrodynamics [16], the Hamilton-Jacobi equation (11) is a well-known Cauchy-Lagrange integral of the Euler equation (4), (6).

The continuity equation (5) taking (9) into account, takes the form

$$\frac{\partial \rho}{\partial t} + \text{div}\left(\frac{1}{m}\rho \nabla S(t,\mathbf{r})\right) = 0 \tag{12}$$

After solving the equation (11), the trajectory of particle motion in the Hamilton-Jacobi theory can be found in two ways: (i) using the Jacobi theorem [1, 2] and the complete integral of the Hamilton-Jacobi equation (11), or (ii) by solving the system of ordinary differential equations

$$\frac{d\mathbf{r}}{dt} = \frac{1}{m}\nabla S(t,\mathbf{r}) \tag{13}$$

It is shown in the Appendix that, for other more complicated Hamiltonian systems, the equation (4) leads to the corresponding Hamilton-Jacobi equations, which allows us to state that the Euler equation (4) is only another form of the corresponding Hamilton-Jacobi equation and the theory (4), (5) is a generalized Hamilton-Jacobi theory describing the Hamilton-Jacobi ensemble.

### 4. Nonrelativistic motion of a particle in a potential field with damping

In this case, the force acting on the particle is

$$\mathbf{F}(t,\mathbf{r},\mathbf{v}) = -\nabla U - \beta \mathbf{v} \tag{14}$$

where $\beta$ is the constant drag factor.

This system in non-Hamoltonian.

Substituting (14) into equation (4), one obtains taking (7) into account

$$m\left(\frac{\partial \mathbf{v}}{\partial t} - [\mathbf{v}\,\text{rot}\,\mathbf{v}]\right) = -\nabla\left(\frac{1}{2}m\mathbf{v}^2 + U\right) - \beta \mathbf{v} \tag{15}$$

This equation also has a solution in the form of a potential flow of the Hamilton-Jacobi ensemble (9), herewith the function $S(t,\mathbf{r})$ satisfies the Hamilton-Jacobi equation

$$\frac{\partial S}{\partial t} + \frac{1}{2m}|\nabla S|^2 + \frac{\beta}{m}S + U(t,\mathbf{r}) = 0 \tag{16}$$

The continuity equation has the form (12).

Another example of the derivation of the Hamilton-Jacobi equation for the non-Hamiltonian system from theory (4), (5) is considered in [15].

We note, however, that for majority of the non-Hamiltonian systems, it is impossible to obtain the Hamilton-Jacobi equation; for such systems the Hamilton-Jacobi theory is formulated in the form of the continual equations (4), (5).

## 5. Generalization of the Hamilton-Jacobi theory

Let us generalize the Hamilton-Jacobi theory considered above to arbitrary dynamical (not necessarily mechanical) systems.

Let there is an arbitrary system whose state at each time is characterized by the generalized coordinates $q = (q_1,..., q_s)$, and its evolution is given by the functions $q_\alpha(t)$, where $\alpha = 1,...,s$; $s$ is the number of degrees of freedom of the system.

Let the evolution (motion) of the system under consideration is described by equations

$$\frac{dp_\alpha}{dt} = F_\alpha(t,q,p) \tag{17}$$

$$\frac{dq_\alpha}{dt} = \varphi_\alpha(t,q,p) \tag{18}$$

where $p = (p_1,..., p_s)$ are the generalized momenta of the system; $v = (v_1,..., v_s)$ are the generalized velocities of the system;

$$v_\alpha = \varphi_\alpha(t,q,p) \tag{19}$$

are the relations connecting the generalized momenta and generalized velocities of the system; $F_\alpha(t,q,v)$ are the generalized forces; $\varphi_\alpha(t,q,p)$ and $F_\alpha(t,q,p)$ are some sufficiently arbitrary algebraic functions.

As above, we introduce for system (17), (18) a $s$-dimensional configuration space in which the state of the system at a given instant of time is represented by a point, while the evolution of the system is represented by a line.

Let us consider the Hamilton-Jacobi ensemble consisting of a plurality of identical systems that differ in the initial conditions and satisfy the condition of single-valuedness: all the systems entering the ensemble that are at the same time in the same point of the configuration space have the same generalized velocity $v$.

In the configuration space, a set of points (the "Hamilton-Jacobi gas") moving in accordance with the equations (17), (18) corresponds to the Hamilton-Jacobi ensemble. The Hamilton-Jacobi ensemble, as a continuous medium, is characterized by the velocity field $v_\alpha(t,q)$ and the density $\rho(t,q)$ at each point $q$ of the configuration space at each instant of time $t$. The transition to the continual (Eulerian) description of the Hamilton-Jacobi ensemble occurs using the relation

$$\frac{d}{dt} = \frac{\partial}{\partial t} + \sum_{\beta=1}^{s} v_\beta \frac{\partial}{\partial q_\beta} \qquad (20)$$

As a result, one obtains the equation of motion and the continuity equation for the Hamilton-Jacobi ensemble in the Eulerian continual approach

$$\frac{\partial p_\alpha}{\partial t} + \sum_{\beta=1}^{s} v_\beta \frac{\partial p_\alpha}{\partial q_\beta} = F_\alpha(t,q,p) \qquad (21)$$

$$\frac{\partial \rho}{\partial t} + \sum_{\alpha=1}^{s} \frac{\partial \rho v_\alpha}{\partial q_\alpha} = 0 \qquad (22)$$

$$v_\alpha = \varphi_\alpha(t,q,p) \qquad (23)$$

The system of equations (21) - (23) is a formulation of the Hamilton-Jacobi theory for an arbitrary system (17), (18).

Let us prove that the system of equations (21) for arbitrary Hamiltonian systems is equivalent to the Hamilton-Jacobi equation.

Let there is a Hamiltonian function $H(t,q,p)$ such that the equations of motion (17), (18) can be written in the form

$$\dot{q}_\alpha = \frac{\partial H}{\partial p_\alpha}, \quad \dot{p}_\alpha = -\frac{\partial H}{\partial q_\alpha} \qquad (24)$$

A system (not necessarily physical) which is described by the equations (24) is the Hamiltonian one, regardless of its nature.

Comparing (24) with (17), (18), we come to the conclusion that for a Hamiltonian system, the generalized force is equal to $F_\alpha(t,q,p) = -\frac{\partial H}{\partial q_\alpha}$, while the generalized velocity is equal to $\varphi_\alpha(t,q,p) = \frac{\partial H}{\partial p_\alpha}$.

Substituting (24) into equation (21), one obtains

$$\frac{\partial p_\alpha}{\partial t} + \sum_{\beta=1}^{s} v_\beta \frac{\partial p_\alpha}{\partial q_\beta} = -\frac{\partial H}{\partial q_\alpha} \qquad (25)$$

Formally, the solutions of equations (21) and (23) have the form $p = p(t,q)$. For this reason, the Hamiltonian function calculated for the Hamilton-Jacobi ensemble will be a function of time and generalized coordinates:

$$\tilde{H}(t,q) = H(t,q,p(t,q)) \qquad (26)$$

Consequently,

$$\frac{\partial \tilde{H}}{\partial q_\alpha} = \frac{\partial H}{\partial q_\alpha} + \sum_{\beta=1}^{s} \frac{\partial H}{\partial p_\beta} \frac{\partial p_\beta}{\partial q_\alpha} \quad (27)$$

Taking into account (24) and (27), we rewrite (25) in the form

$$\frac{\partial p_\alpha}{\partial t} + \frac{\partial \tilde{H}}{\partial q_\alpha} + \sum_{\beta=1}^{s} v_\beta \left( \frac{\partial p_\alpha}{\partial q_\beta} - \frac{\partial p_\beta}{\partial q_\alpha} \right) = 0 \quad (28)$$

Equation (28) has a potential solution

$$p_\alpha = \frac{\partial S}{\partial q_\alpha} \quad (29)$$

where the potential function $S(t,q)$ satisfies the equation

$$\frac{\partial S}{\partial t} + \tilde{H}(t,q) = 0$$

which taking into account (26) and (29) is the Hamilton-Jacobi equation

$$\frac{\partial S}{\partial t} + H\left(t, q, \frac{\partial S}{\partial q}\right) = 0 \quad (30)$$

The continuity equation (22) in this case takes the form

$$\frac{\partial \rho}{\partial t} + \sum_{\alpha=1}^{s} \frac{\partial}{\partial q_\alpha} \left[ \rho \varphi_\alpha \left(t, q, \frac{\partial S}{\partial q}\right) \right] = 0 \quad (31)$$

Thus, we have proved that for arbitrary Hamiltonian systems (24) the equation of motion (21), (23) is equivalent to the Hamilton-Jacobi equation (30). For this reason, we can say that the system of equations (21) - (23) is a generalization of the Hamilton-Jacobi theory to an arbitrary system described by the equations of motion (17), (18). In the case of non-Hamiltonian systems, the vector equation (21), (23) may not have a potential solution (29) and therefore it cannot always be reduced to one scalar Hamilton-Jacobi equation.

## 6. Hamilton-Jacobi theory in $p$-representation

Equations (17) and (18) are formally symmetric with respect to the generalized momenta $p$ and generalized coordinates $q$. This allows us to formulate the Hamilton-Jacobi theory in momentum space.

Let us consider the momentum space $p = (p_1, ..., p_s)$. The system at each instant of time can be represented as a point in momentum space, herewith the evolution of the system is described by a line in momentum space, while the velocity of the system in momentum space, according to (17), is

$$\omega_\alpha = F_\alpha(t, q, p) \tag{32}$$

Let us consider a set of identical systems differing by initial conditions, but satisfying the single-valuedness requirement: all systems having at the same time the same generalized momenta $p$ must have the same generalized coordinates $q$. Obviously, such an ensemble is the Hamilton-Jacobi ensemble defined above.

The Hamilton-Jacobi ensemble is represented by a set of points moving in momentum space according to equations (17), (18), herewith the trajectories of these points in the momentum space do not intersect. In this case, equation (17) can be considered as the determination of the velocity of the system in momentum space, while the equation (18) can be considered as the equation of motion which plays a role of the Newton's equation in momentum space.

Considering the movement of the ensemble in momentum space, we introduce the material time derivative $\dfrac{d}{dt}$ which characterizes the changes associated with a specific system of the ensemble and the local time derivative $\dfrac{\partial}{\partial t}$ which describes the change in the parameters of the ensemble at a fixed point in the momentum space.

These derivatives are related by the obvious relation

$$\frac{d}{dt} = \frac{\partial}{\partial t} + \sum_{\alpha=1}^{s} \omega_\alpha \frac{\partial}{\partial p_\alpha} \tag{33}$$

According to the single-valuedness requirement, one can introduce a vector function $q(t, p)$ defined in momentum space, which describes the generalized coordinates of the system which is at a given instant of time at a given point in the momentum space.

In this case, the equation of motion (18), taking into account (33), can be rewritten in Eulerian form

$$\frac{\partial q_\alpha}{\partial t} + \sum_{\beta=1}^{s} \omega_\beta \frac{\partial q_\alpha}{\partial p_\beta} = \varphi_\alpha(t, q, p) \tag{34}$$

where the velocities $\omega$ of the Hamilton-Jacobi ensemble in momentum space are determined by the relation (32).

Solving equations (34), one can obtain, at least formally, the dependences $q(t, p)$ that allow finding the velocities of the Hamilton-Jacobi ensemble at each point of the momentum space

$$\omega_\alpha(t, p) = F_\alpha(t, q(t, p), p) \tag{35}$$

Let us introduce the density of the Hamilton-Jacobi ensemble in momentum space $\rho(t, p)$, as the number of systems per unit volume of momentum space. Obviously, this density satisfies the continuity equation

$$\frac{\partial \rho}{\partial t} + \sum_{\alpha=1}^{s} \frac{\partial \rho \omega_\alpha}{\partial p_\alpha} = 0 \qquad (36)$$

where $\omega(t, p)$ is defined by the relation (35).

Equations (32), (34), (36) constitute the Hamilton-Jacobi theory in the *p*-representation.

As an example, let us consider a system consisting of a single particle moving under the action of a force $\mathbf{F}(t, \mathbf{r}, \mathbf{v})$. In the *p*-representation, the Hamilton-Jacobi ensemble for such a system is described by the functions $\mathbf{r} = \mathbf{r}(t, \mathbf{v})$ and $\rho = \rho(t, \mathbf{v})$ that satisfy the equations

$$\frac{\partial \mathbf{r}}{\partial t} + \frac{1}{m}\left(\mathbf{F}\frac{\partial}{\partial \mathbf{v}}\right)\mathbf{r} = \mathbf{v} \qquad (37)$$

$$\frac{\partial \rho}{\partial t} + \frac{1}{m}\frac{\partial \rho \mathbf{F}}{\partial \mathbf{v}} = 0 \qquad (38)$$

Equation (37), or in the general form, the equation (34), is a new formulation of classical mechanics in the *p*-representation.

Let us consider in the general form the *p*-representation for the Hamiltonian system (24). In this case, the equation (34) takes the form

$$\frac{\partial q_\alpha}{\partial t} + \sum_{\beta=1}^{s} \omega_\beta \frac{\partial q_\alpha}{\partial p_\beta} = \frac{\partial H}{\partial p_\alpha} \qquad (39)$$

For the Hamilton-Jacobi ensemble we introduce

$$\hat{H}(t, p) = H(t, q(t, p), p) \qquad (40)$$

Obviously,

$$\frac{\partial \hat{H}}{\partial p_\alpha} = \frac{\partial H}{\partial p_\alpha} + \sum_{\beta=1}^{s} \frac{\partial H}{\partial q_\beta}\frac{\partial q_\beta}{\partial p_\alpha}$$

Then the equation (39) can be rewritten in the form

$$\frac{\partial q_\alpha}{\partial t} + \sum_{\beta=1}^{s} \omega_\beta \left(\frac{\partial q_\alpha}{\partial p_\beta} - \frac{\partial q_\beta}{\partial p_\beta}\right) = \frac{\partial \hat{H}}{\partial p_\alpha}$$

This equation has a potential solution

$$q_\alpha = -\frac{\partial \Phi}{\partial p_\alpha} \qquad (41)$$

where the potential $\Phi(t, p)$ satisfies the equation

$$\frac{\partial \Phi}{\partial t} + \hat{H}(t, p) = 0$$

which together with (40) and (41) is the Hamilton-Jacobi equation in the momentum representation

$$\frac{\partial \Phi}{\partial t} + H\left(t, -\frac{\partial \Phi}{\partial p}, p\right) = 0 \qquad (42)$$

The continuity equation (36) in the *p*-representation takes the form

$$\frac{\partial \rho}{\partial t} + \sum_{\alpha=1}^{s} \frac{\partial}{\partial p_\alpha}\left[\rho F_\alpha\left(t, -\frac{\partial \Phi}{\partial p}, p\right)\right] = 0 \qquad (43)$$

If the Hamiltonian function does not depend explicitly on time, the equation (42) has a stationary solution

$$\Phi(t, p) = -Et + W(p) \qquad (44)$$

where the function $W(p)$ satisfies the truncated Hamilton-Jacobi equation in the *p*-representation

$$H\left(-\frac{\partial W}{\partial p}, p\right) = E \qquad (45)$$

The problem of the motion of a Hamiltonian system can be solved using the Jacobi theorem and the complete integral of the Hamilton-Jacobi equation (41), similarly to the Hamilton-Jacobi equation (30) in the $q$-representation [1]. The Jacobi theorem in the **p**-representation is easily proved using canonical transformations [1]. Here, for brevity, we give the Jacobi theorem without proof. Let the complete integral of the Hamilton-Jacobi equation (42) is known in the form

$$\Phi = \Phi(t, p_1,..., p_s; \beta_1,...\beta_s) + A$$

where $\beta_1,...\beta_s, A$ are the arbitrary constants.

Differentiating this function with respect to arbitrary constants $\beta$ and equating derivatives to the new constants $\alpha$, one obtains a system of algebraic equations

$$\frac{\partial \Phi}{\partial \beta_i} = \alpha_i$$

solving which, we can find the momenta $p$, as the functions of time and $2s$ arbitrary constants. The dependence of the coordinates on time, then, is found from equations (41).

The connection between the Hamilton-Jacobi theory in the *p*-representation (42), (43) with the formulation of quantum mechanics in the momentum representation [17] is obvious.

## 7. Topology of the configuration space

The configuration space of the classical Hamilton-Jacobi ensemble can have a complex topology. In particular, the regions that are not accessible for the motion of classical systems can exist in the configuration space. In this case, the motion occurs only within the "accessible" regions that are separated from the "forbidden" areas by some hypersurfaces $\Gamma_\alpha$, where $\alpha = 1,2,...$ is the number of the boundary hypersurface. The configuration space can be either simply connected or multiply connected (non-simply connected). In the case of a multiply connected configuration space, the regions accessible for the motion of a classical system are separated from one another by the "forbidden" regions. In this case, the movement of the classical system can occur only within one of the accessible regions, and the system cannot spontaneously jump from one accessible region to another through the forbidden region. A simple example is the one-dimensional motion of a particle in double well potential, whose height exceeds the energy of a particle.

According to the definition of the Hamilton-Jacobi ensemble, all the particles that are at a given point in the configuration space at a given moment of time must have the same velocity. It is easy to see that for some systems (in particular for systems that perform finite motions) this condition is not satisfied for the entire time of the motion of the system. For example, a harmonic oscillator can travel through the same point of space with different (in direction) velocities. In this case, all harmonic oscillators having the same energy but moving in the forward and reverse directions cannot be assigned to the same Hamilton-Jacobi ensemble.

The same applies to other more complex systems that perform finite motions. From a mathematical point of view, this means that equations (21) - (23) (or (30), (31) for Hamiltonian systems) have several solutions in the same region of the configuration space. We denote these solutions as $(\rho^{(n)}, v^{(n)})$ or for Hamiltonian systems as $(\rho^{(n)}, S^{(n)})$, where $n = 1,2,...$ is the number of solution. Such a non-uniqueness of the solution of equations (21) - (23) is inconsistent with the definition of the Hamilton-Jacobi ensemble.

This difficulty can be avoided if we assume that the configuration space of the system consists of several "parallel layers" (sheets) glued along the boundary hypersurfaces $\Gamma_\alpha$. Such a space is similar to the Riemann surfaces for many-valued functions of a complex variable. It is assumed that on each of these layers of a multilayer configuration space, only one solution from the set of allowed solutions of equations (21) - (23) is realized. For each partial solution $n$, the functions $\rho^{(n)}(t,q), v^{(n)}(t,q)$ and, respectively, $S^{(n)}(t,q)$ (for Hamiltonian systems) are single-valued. Thus, on each layer of a multilayer configuration space, a partial Hamilton-Jacobi ensemble can be defined that satisfies the single-valuedness condition.

If the system is on some layer $n$ of a multilayer configuration space and in its movement reached a point in which at least one component of the velocity of the system changes its sign (at this point, this component of the velocity is zero), then the system proceeds to another layer of the multilayer configuration space and continues its movement already on it. The set of points at which the system passes from one layer to another form the hypersurfaces on which the layers of the multilayer configuration space "glue" together. It is easy to see that the system cannot go beyond the region bounded by these hypersurfaces, because the vector of velocity of the system is directed along the tangent of the hypersurfaces. From this, we conclude that the layers of the multilayer configuration space are glued together over the hypersurfaces $\Gamma_\alpha$ that limit the regions of the configuration space accessible for the classical motion. Thus, when a hypersurface $\Gamma_\alpha$ is reached, the classical system passes to another layer of the multilayer configuration space and continues its motion on it.

From the condition of continuity of the motion of the Hamilton-Jacobi ensemble, we can obtain the condition on hypersurfaces $\Gamma_\alpha$

$$j_\perp^{(n)}(t,\Gamma_\alpha) = -j_\perp^{(k)}(t,\Gamma_\alpha)$$

which reflects the fact that the number of systems of an ensemble that have reached the hypersurface $\Gamma_\alpha$ when moving along a layer $n$ is equal to the number of systems that have passed to another layer $k$, where

$$j^{(n)} = \rho^{(n)}(t,q)v^{(n)}(t,q)$$

is the flux density of the Hamilton-Jacobi ensemble on the layer $n$ of the configuration space; index $\perp$ means the projection of the vector $j$ onto the outer normal to the hypersurface $\Gamma_\alpha$.

Note that if the observation of the movement of the classical system does not proceed continuously, but only discrete events of its interaction with a measuring device are detected (this is precisely what happens when observing quantum objects), then the density of the ensemble $\rho$ with appropriate normalization can be interpreted as the probability density to detect the system at a given point in the configuration space.

Since the system can interact with the measuring device, being at a given point in the configuration space during any admissible partial motion, the total probability density to detect the system at a given point in the configuration space will be determined by the relation

$$\rho = \sum_n w_n \rho^{(n)} \tag{46}$$

where $w_n$ is the probability of finding the $n$-th partial Hamilton-Jacobi ensemble in the total flow of systems.

Obviously,

$$\sum_n w_n = 1 \tag{47}$$

Then the total flux of all systems at a given point in the configuration space, no matter what partial motion they make, is determined by the obvious relation

$$j = \sum_n w_n \rho^{(n)} v^{(n)} \qquad (48)$$

In the case of quantum Hamiltonian systems, the Schrödinger equation is linear and its solution is a superposition of partial solutions $\psi^{(n)}(t,q)$:

$$\psi(t,q) = \sum_n a_n \psi^{(n)}(t,q) \qquad (49)$$

According to the probabilistic interpretation of quantum mechanics [4]

$$|a_n|^2 = w_n \qquad (50)$$

In WKB approximation, the partial solutions of the Schrödinger equation $\psi^{(n)}(t,q)$ can be represented in the form [4]

$$\psi^{(n)}(t,q) = \sqrt{\rho^{(n)}(t,q)} \exp\left(i \frac{S^{(n)}(t,q)}{\hbar}\right) \qquad (51)$$

where the functions $\rho^{(n)}(t,q)$ and $S^{(n)}(t,q)$ correspond to the partial solutions of equations (30), (31) for the corresponding classical Hamiltonian system, i.e. to the partial Hamilton-Jacobi ensembles. Thus, the wave functions $\psi^{(n)}(t,q)$ that are the partial solutions, for example, of Schrödinger equation, describe the partial quantum Hamilton-Jacobi ensembles that perform unidirectional motions in the configuration space. Continuing the analogy with the classical Hamilton-Jacobi ensembles, we can say that the partial wave functions (51) are defined on the different layers of a multilayer configuration space, while the result of a measurement carried out over the system is their interference (49). If desired, in this one can be find an analogy with the many-worlds interpretation of quantum mechanics [5,6], although, obviously, that the Hamilton-Jacobi ensemble and the multilayer configuration space are nothing more than the convenient abstractions.

The fundamental difference between quantum systems and classical ones is the interference of the states (49), which manifests itself in the fact that not the densities of the partial Hamilton-Jacobi ensembles are summed up as in (46), but the wave functions (51) corresponding to different partial Hamilton-Jacobi ensembles (i.e. corresponding to the movements of the system on the different layers of the multilayer configuration space).

We note that the total wave function of the system (49), which can formally also be represented in the form (51), does not correspond to any solution of the Hamilton-Jacobi equation for classical system. This means that there is no the Hamilton-Jacobi ensemble which corresponds to the wave function (49). The exception is "unidirectional" motion, when there is a unique partial Hamilton-Jacobi ensemble described by a single-valued solution of the Hamilton-Jacobi equation and, accordingly, by the unique solution of the Schrödinger equation. This is possible only for infinite motion of the system.

Thus we conclude that only partial waves in quantum mechanics can correspond to the partial Hamilton-Jacobi ensembles in classical mechanics.

This allows clarifying the transition from quantum mechanics to classical mechanics in the Hamilton-Jacobi form: in the limit $\hbar \to 0$, the Schrödinger equation turns into the Hamilton-Jacobi theory for the corresponding classical Hamiltonian system only as applied to each Hamilton-Jacobi partial ensemble separately.

Hence, in particular, it follows that the so-called Bohmian mechanics [7-9] is applicable only to the "unidirectional" motion, or in the case of complex motions described by standing waves, to each partial Hamilton-Jacobi ensemble separately, and it is incapable of describing the total motion of quantum particle, which is described by the superposition of partial waves (49).

## 8. Concluding remarks

The analysis above provides a general method for constructing the continuum theory of the Hamilton-Jacobi ensemble, and in those cases, where this is possible, gives a physically clear method for deriving the Hamilton-Jacobi equations. Considering quantum mechanics as an ensemble theory, the proposed method gives a visual transition from classical mechanics to quantum mechanics and back.

Obviously, in order to be able to construct the Hamilton-Jacobi theory in the generalized sense, the system must satisfy the locality condition: all functions characterizing the system must depend only on the current parameters of the system. In particular, the forces should depend only on the current generalized coordinates and the current generalized momenta (velocities) of the system, while the generalized momenta should be uniquely expressed in terms of the current velocities and the current coordinates of the system.

Hence, in particular, it follows that it is impossible to construct a relativistic Hamilton-Jacobi theory for a system of interacting charged particles, neither in the form of continual equations of motion, nor even more so in the form of the relativistic Hamilton-Jacobi equation. This is due to the finite speed of propagation of the electromagnetic interaction and the retarded potentials that arise from this, which depend not only on the current space-time coordinates of the particles, but also on the entire prehistory of their motion.

This, in turn, leads to the inability to construct a relativistic many-particle wave equation for a system of interacting charged quantum particles. For this reason, within the framework of the present approach, the many-particle relativistic quantum system can be described only using the approximate methods, such as the method of second quantization.

## APPENDIX

### A1. Nonrelativistic particle in an electromagnetic field

The Lorentz force $\mathbf{F} = e\left(\mathbf{E} + \frac{1}{c}[\mathbf{vH}]\right)$ acts on the particle, where $e$ is the particle charge, $\mathbf{E} = -\nabla\varphi - \frac{1}{c}\frac{\partial \mathbf{A}}{\partial t}, \mathbf{H} = \text{rot}\mathbf{A}$ are the electric and magnetic field strengths, $\varphi, \mathbf{A}$ are the scalar and vector potentials of the electromagnetic field.

The Euler equation (4) for the Hamilton-Jacobi ensemble has the form

$$m\left(\frac{\partial \mathbf{v}}{\partial t} + (\mathbf{v}\nabla)\mathbf{v}\right) = e\left(\mathbf{E} + \frac{1}{c}[\mathbf{vH}]\right) \tag{A1.1}$$

Substituting in (A1.1) the strengths of the electric and magnetic fields, expressed in terms of the scalar and vector potentials of the electromagnetic field, one obtains

$$\frac{\partial}{\partial t}\left(m\mathbf{v} + \frac{e}{c}\mathbf{A}\right) + m(\mathbf{v}\nabla)\mathbf{v} - \frac{e}{c}[\mathbf{v}\text{rot}\mathbf{A}] = -e\nabla\varphi \tag{A1.2}$$

Taking (7) into account, we can rewrite the equation (A1.2) in the form

$$\frac{\partial}{\partial t}\left(m\mathbf{v} + \frac{e}{c}\mathbf{A}\right) + \nabla\left(\frac{mV^2}{2} + e\varphi\right) = [\mathbf{v}\,\text{rot}\left(m\mathbf{v} + \frac{e}{c}\mathbf{A}\right)] \tag{A1.3}$$

This equation has a solution

$$m\mathbf{v} + \frac{e}{c}\mathbf{A} = \nabla S \tag{A1.4}$$

where the function $S$ satisfies the equation

$$\frac{\partial S}{\partial t} + \frac{m}{2}\mathbf{v}^2 + e\varphi = 0$$

which taking into account (A1.4) is the Hamilton-Jacobi equation for nonrelativistic particles in external electromagnetic field [18]

$$\frac{\partial S}{\partial t} + \frac{1}{2m}\left(\nabla S - \frac{e}{c}\mathbf{A}\right)^2 + e\varphi = 0 \tag{A1.5}$$

Correspondingly, the continuity equation (5) takes the form

$$\frac{\partial \rho}{\partial t} + \text{div}\left[\rho \frac{1}{m}\left(\nabla S - \frac{e}{c}\mathbf{A}\right)\right] = 0 \tag{A1.6}$$

## A2. Nonrelativistic particle in an electromagnetic field in the presence of damping

Suppose that a drag force $\mathbf{f}$ which linearly depends on the particle velocity acts on the particle, in addition to the Lorentz force.

We first consider the case

$$\mathbf{f} = -\beta \mathbf{v} \tag{A2.1}$$

The total force acting on a particle is

$$\mathbf{F} = e\left(\mathbf{E} + \frac{1}{c}[\mathbf{vH}]\right) - \beta \mathbf{v} \tag{A2.2}$$

By analogy with Appendix A1, we can reduce the Euler's equation to the form

$$\frac{\partial}{\partial t}\left(m\mathbf{v} + \frac{e}{c}\mathbf{A}\right) + \nabla\left(\frac{mV^2}{2} + e\varphi\right) = [\mathbf{v}\,\text{rot}\left(m\mathbf{v} + \frac{e}{c}\mathbf{A}\right)] - \beta \mathbf{v}$$

This equation does not have a potential solution in the general case, and therefore the Hamilton-Jacobi equation cannot be derived for such a system.

It was assumed above that the drag force is described by the expression (A2.1), regardless of whether there is an external magnetic field or not. However, in the presence of a magnetic field, another generalization of the drag force (A2.1) is possible:

$$\mathbf{f} = -\frac{\beta}{m}\left(m\mathbf{v} + \frac{e}{c}\mathbf{A}\right) \tag{A2.3}$$

In this case, the force acting on the particle is

$$\mathbf{F} = e\left(\mathbf{E} + \frac{1}{c}[\mathbf{vH}]\right) - \frac{\beta}{m}\left(m\mathbf{v} + \frac{e}{c}\mathbf{A}\right) \tag{A2.4}$$

and the Euler's equation for the Hamilton-Jacobi ensemble takes the form

$$\frac{\partial}{\partial t}\left(m\mathbf{v} + \frac{e}{c}\mathbf{A}\right) + \nabla\left(\frac{mV^2}{2} + e\varphi\right) = [\mathbf{v}\,\text{rot}\left(m\mathbf{v} + \frac{e}{c}\mathbf{A}\right)] - \frac{\beta}{m}\left(m\mathbf{v} + \frac{e}{c}\mathbf{A}\right)$$

This equation has a potential solution

$$m\mathbf{v} + \frac{e}{c}\mathbf{A} = \nabla S \tag{A2.5}$$

where the function $S$ satisfies the Hamilton-Jacobi equation

$$\frac{\partial S}{\partial t} + \frac{1}{2m}(\nabla S - \frac{e}{c}\mathbf{A})^2 + e\varphi + \frac{\beta}{m}S = 0 \tag{A2.6}$$

The continuity equation for the Hamilton-Jacobi ensemble has the form (A1.6).

### A3. Nonrelativistic magnetic dipole in an electromagnetic field

In paper [12], the Hamilton-Jacobi theory for the classical magnetic dipole was constructed. Let us present the derivation of this theory using the approach developed in this paper.

The nonrelativistic motion of a classical magnetic dipole in an electromagnetic field is described by a system of equations

$$m\frac{d\mathbf{v}}{dt} = e\left(\mathbf{E} + \frac{1}{c}\mathbf{v}\times\mathbf{H}\right) + \gamma(\mathbf{s}\nabla)\mathbf{H} \tag{A3.1}$$

$$\frac{d\mathbf{s}}{dt} = \gamma\mathbf{s}\times\mathbf{H} \tag{A3.2}$$

$$\frac{d\mathbf{r}}{dt} = \mathbf{v} \tag{A3.3}$$

where $\gamma$ is the internal gyromagnetic ratio of the magnetic dipole.

According to equation (A3.2), the angular momentum of a particle $\mathbf{s}$ has a constant modulus, which, for example, for an electron, is assumed to be equal to $\frac{1}{2}\hbar$. This means that one can introduce two angles $\theta$ and $\chi$ of a spherical coordinate system that completely determine the orientation of the vector $\mathbf{s}$ in space. In this case, the vector $\mathbf{s}$ can be represented in the form

$$\mathbf{s} = \frac{1}{2}\hbar(\sin\theta\sin\chi, \sin\theta\cos\chi, \cos\theta) \tag{A3.4}$$

Instead of an angle $\theta$, we use the coordinate $\xi$, which is defined by

$$\xi = \frac{1}{2}\hbar\cos\theta \tag{A3.5}$$

Obviously,

$$\xi = s_z \tag{A3.6}$$

As shown in [12,19], for the system (A3.1) - (A3.3) one can introduce the Hamiltonian

$$H(\mathbf{P},\mathbf{r},\xi,\chi) = \frac{1}{2m}\left(\mathbf{P} - \frac{e}{c}\mathbf{A}\right)^2 + e\varphi - \gamma(\mathbf{sH}) \tag{A3.7}$$

where

$$\mathbf{P} = m\mathbf{v} + \frac{e}{c}\mathbf{A} \tag{A3.8}$$

is the generalized momentum; the parameter

$$H_{sp} = -\gamma(\mathbf{sH}) \tag{A3.9}$$

is a component of the Hamiltonian associated with the interaction of the intrinsic magnetic moment of the classical particle with an external magnetic field.

Then the equations of motion of the classical magnetic dipole (A3.1) - (A3.3) can be written in the form of Hamilton's equations [12, 19]

$$\frac{d\mathbf{P}}{dt} = -\frac{\partial H}{\partial \mathbf{r}} \qquad (A3.10)$$

$$\frac{d\mathbf{r}}{dt} = \frac{\partial H}{\partial \mathbf{P}} \qquad (A3.11)$$

$$\frac{d\chi}{dt} = -\frac{\partial H}{\partial \xi} \qquad (A3.12)$$

$$\frac{d\xi}{dt} = \frac{\partial H}{\partial \chi} \qquad (A3.13)$$

Let us consider the Hamilton-Jacobi ensemble in the configuration space $\mathbf{r}$.

This ensemble is described by the fields $\mathbf{v}(t,\mathbf{r})$, $\xi(t,\mathbf{r})$ and $\chi(t,\mathbf{r})$. The transition from the Lagrangian description to the Eulerian description occurs by means of the formal replacement in equations (A3.10), (A3.12) and (A3.13) of the material derivative with respect to time by the relation (3).

As a result, equations (A3.10), (A3.12) and (A3.13) take the form

$$\frac{\partial \mathbf{P}}{\partial t} + (\mathbf{v}\nabla)\mathbf{P} = -\nabla H + \left(\frac{\partial H}{\partial P_i}\right)_{\mathbf{r},\xi,\chi} \nabla P_i + \left(\frac{\partial H}{\partial \xi}\right)_{\mathbf{P},\mathbf{r},\chi} \nabla \xi + \left(\frac{\partial H}{\partial \omega}\right)_{\mathbf{P},\mathbf{r},\xi} \nabla \chi \qquad (A3.14)$$

$$\frac{\partial \chi}{\partial t} + (\mathbf{v}\nabla)\chi = -\left(\frac{\partial H}{\partial \xi}\right)_{\mathbf{P},\mathbf{r},\chi} \qquad (A3.15)$$

$$\frac{\partial \xi}{\partial t} + (\mathbf{v}\nabla)\xi = \left(\frac{\partial H}{\partial \chi}\right)_{\mathbf{P},\mathbf{r},\xi} \qquad (A3.16)$$

where $i = 1,2,3$; the summation is performed over the repeated indices.

Equation (A3.11) taking into account (3) and (A3.8) turns into an identity.

The lower indices indicate the parameters considered in differentiation as constants.

Taking into account (A3.11), (A3.15) and (A3.16), one rewrites (A3.14) in the form

$$\frac{\partial \mathbf{P}}{\partial t} + (\mathbf{v}\nabla)\mathbf{P} = -\nabla H + v_i \nabla P_i - \frac{\partial \chi}{\partial t} \nabla \xi - \nabla \xi (\mathbf{v}\nabla)\chi + \frac{\partial \xi}{\partial t} \nabla \chi + \nabla \chi (\mathbf{v}\nabla)\xi \qquad (A3.17)$$

or after simple transforms

$$\frac{\partial}{\partial t}(\mathbf{P} - \xi\nabla\chi) + (\mathbf{v}\nabla)(\mathbf{P} - \xi\nabla\chi) = -\nabla(\xi\frac{\partial \chi}{\partial t} + H) + v_i \nabla(P_i - \xi\partial_i\chi) \qquad (A3.18)$$

where $\partial_i = \partial/\partial x_i$.

Taking into account the identity

$$v_i \nabla (P_i - \xi \partial_i \chi) - (\mathbf{v}\nabla)(\mathbf{P} - \xi \nabla \chi) = \mathbf{v} \times \nabla \times (\mathbf{P} - \xi \nabla \chi) \qquad (A3.19)$$

we can rewrite (A3.18) in the form

$$\frac{\partial}{\partial t}(\mathbf{P} - \xi \nabla \chi) = -\nabla(\xi \frac{\partial \chi}{\partial t} + H) + \mathbf{v} \times \nabla \times (\mathbf{P} - \xi \nabla \chi) \qquad (A3.20)$$

This equation has a solution in the form

$$\mathbf{P} - \xi \nabla \chi = \nabla S \qquad (A3.21)$$

where the function $S(t,\mathbf{r})$ satisfies the equation

$$\frac{\partial S}{\partial t} + \xi \frac{\partial \chi}{\partial t} + H = 0 \qquad (A3.22)$$

From (A3.8) and (A3.21) it follows

$$m\mathbf{v} = \nabla S - \frac{e}{c}\mathbf{A} + \xi \nabla \chi \qquad (A3.23)$$

Thus, the Hamilton-Jacobi ensemble for classical magnetic dipoles is described by the equations [12]

$$\frac{\partial S}{\partial t} + \xi \frac{\partial \chi}{\partial t} + \frac{1}{2m}\left(\nabla S - \frac{e}{c}\mathbf{A} + \xi \nabla \chi\right)^2 + e\varphi + H_{sp} = 0 \qquad (A3.24)$$

$$\frac{\partial \chi}{\partial t} + (\mathbf{v}\nabla)\chi = -\frac{\partial H_{sp}}{\partial \xi} \qquad (A3.25)$$

$$\frac{\partial \xi}{\partial t} + (\mathbf{v}\nabla)\xi = \frac{\partial H_{sp}}{\partial \chi} \qquad (A3.26)$$

Herewith, the density $\rho(t,\mathbf{r})$ of the Hamilton-Jacobi ensemble in the configuration space is described by the continuity equation (5), which, taking (A3.23) into account, takes the form

$$\frac{\partial \rho}{\partial t} + \nabla\left(\rho(\nabla S - \frac{e}{c}\mathbf{A} + \xi \nabla \chi)\right) = 0 \qquad (A3.27)$$

Equations (A3.24) - (A3.27) constitute the Hamilton-Jacobi theory for a nonrelativistic magnetic dipole.

Let us note that here, when constructing the Hamilton-Jacobi ensemble, we assumed that at a given point in the configuration space at a given time, all the magnetic dipoles have the same velocity and the same orientation of the magnetic moment. But it is easy to construct a theory in which the less strong constraints are imposed on the Hamilton-Jacobi ensemble. To do this, we can expand the definition of the configuration space and consider the 5-dimensional configuration space $(\mathbf{r},\chi,\xi)$.

## A4. The motion of a charged particle in the electromagnetic and gravitational fields

The equation of motion of a charged particle in the electromagnetic and gravitational fields has the form [18]

$$\frac{du^i}{ds} + \Gamma^i{}_{kl} u^k u^l = \frac{e}{mc^2} F^{ik} u_k \tag{A4.1}$$

where $e$ is the electric charge of the particle, $m$ is its mass, $ds^2 = g_{ik} dx^i dx^k$, $i,k = 0,1,2,3$,

$$F_{ik} = \frac{\partial A_k}{\partial x^i} - \frac{\partial A_i}{\partial x^k}$$

is the tensor of the electromagnetic field; $A_i$ is the 4-potential of the electromagnetic field, $u^i = \frac{dx^i}{ds}$ is the 4-velocity of the particle, which by definition satisfies the condition [18]

$$g_{kl} u^k u^l = 1 \tag{A4.2}$$

the metric tensor $g_{kl}$ describes the external gravitational field.

In the covariant form, the equation of motion (A4.1) has the form [18]

$$mc \left( \frac{du_i}{ds} - \Gamma_{i,kl} u^k u^l \right) = \frac{e}{c} F_{ik} u^k$$

Substituting here the Christoffel symbols $\Gamma_{i,kl}$, one obtains [18]

$$mc \left( \frac{du_i}{ds} - \frac{1}{2} \frac{\partial g_{kl}}{\partial x^i} u^k u^l \right) = \frac{e}{c} F_{ik} u^k \tag{A4.3}$$

Let us turn to the Hamilton-Jacobi ensemble. In this case we have velocity fields $u^i(x)$ and can write

$$\frac{du^i}{ds} = u^k \frac{\partial u^i}{\partial x^k}$$

Then taking into account the definition of $F_{ik}$, the equation of motion (A4.3) for the Hamilton-Jacobi ensemble takes the form

$$u^k \left( mc \frac{\partial u_i}{\partial x^k} - mc \frac{1}{2} \frac{\partial g_{kl}}{\partial x^i} u^l - \frac{e}{c} \frac{\partial A_k}{\partial x^i} + \frac{e}{c} \frac{\partial A_i}{\partial x^k} \right) = 0 \tag{A4.4}$$

For the Hamilton-Jacobi ensemble, one can introduce a 4-vector of flux

$$j^i = \rho_0 u^i \tag{A4.5}$$

which satisfies the continuity equation [18]

$$\frac{1}{\sqrt{-g}} \frac{\partial}{\partial x^i}\left(\sqrt{-g}\, j^i\right) = 0 \tag{A4.6}$$

where $\rho_0$ is the 4-scalar.

Equations (A4.4) - (A4.5) describe the motion of the Hamilton-Jacobi ensemble and constitute the Hamilton-Jacobi theory in the generalized sense.

Differentiating (A4.2), one obtains

$$\frac{\partial g_{kl}}{\partial x^i} u^k u^l + 2 g_{kl} \frac{\partial u^l}{\partial x^i} u^k = 0 \tag{A4.7}$$

Taking into account that $u^l = g^{lk} u_k$, one has

$$g_{kl} \frac{\partial u^l}{\partial x^i} = g_{kl} \frac{\partial g^{lm}}{\partial x^i} u_m + \frac{\partial u_k}{\partial x^i} \tag{A4.8}$$

Taking into account (A4.8), the expression (A4.7) takes the form

$$\frac{1}{2} \frac{\partial g_{kl}}{\partial x^i} u^k u^l + g_{kl} \frac{\partial g^{lm}}{\partial x^i} u^k u_m + \frac{\partial u_k}{\partial x^i} u^k = 0 \tag{A4.9}$$

Using (A4.9), equation (A4.4) can be rewritten in the form

$$u^k \left( mc \frac{\partial u_i}{\partial x^k} + \frac{e}{c} \frac{\partial A_i}{\partial x^k} - mc \frac{\partial u_k}{\partial x^i} - \frac{e}{c} \frac{\partial A_k}{\partial x^i} - mc \frac{\partial g_{kl}}{\partial x^i} u^l - mc g_{kl} \frac{\partial g^{lm}}{\partial x^i} u_m \right) = 0 \tag{A4.10}$$

It is easy to prove that

$$\frac{\partial g_{kl}}{\partial x^i} u^l + g_{kl} \frac{\partial g^{lm}}{\partial x^i} u_m = 0 \tag{A4.11}$$

To do this, we use the well-known property of the metric tensor [18]

$$g_{ks} \frac{\partial g^{sm}}{\partial x^i} = -g^{lm} \frac{\partial g_{kl}}{\partial x^i}$$

which can be rewritten in the form

$$\frac{\partial g_{kl}}{\partial x^i} + g_{ks} g_{ml} \frac{\partial g^{sm}}{\partial x^i} = 0$$

Thus, taking into account (A4.11), the equation (A4.10) takes the form

$$u^k \left( \frac{\partial}{\partial x^k}\left( mc u_i + \frac{e}{c} A_i \right) - \frac{\partial}{\partial x^i}\left( mc u_k + \frac{e}{c} A_k \right) \right) = 0 \tag{A4.12}$$

The equation (A4.12) has the solution

$$mc u_i + \frac{e}{c} A_i = -\frac{\partial S}{\partial x^i} \tag{A4.13}$$

which, when substituted in (A4.2), leads to the Hamilton-Jacobi equation

$$g^{ik}\left(\frac{\partial S}{\partial x^i}+\frac{e}{c}A_i\right)\left(\frac{\partial S}{\partial x^k}+\frac{e}{c}A_k\right)=m^2c^2 \tag{A4.14}$$

where $S$ is the 4-scalar.

Taking (A4.13) into account, the continuity equation (A4.6) for the 4-vector of flux (A4.5) can be written in the form

$$\frac{1}{\sqrt{-g}}\frac{\partial}{\partial x^i}\left(\sqrt{-g}\,\rho_0 g^{ik}\left(\frac{\partial S}{\partial x^k}+\frac{e}{c}A_k\right)\right)=0 \tag{A4.15}$$

Equations (A4.14) and (A4.15) constitute the Hamilton-Jacobi theory in the narrow sense for a charged particle moving in an electromagnetic field, both in curved and in flat space.

### A.5. Nonrelativistic system of interacting particles

Let us consider a system consisting of $N$ interacting particles, which in general case have different masses $m_i$, $i=1,...,N$.

Equations of motion of the particles have the form

$$m_i\frac{d\mathbf{v}_i}{dt}=\mathbf{F}_i(t,\mathbf{r}_1,...\mathbf{r}_N,\mathbf{v}_1,...,\mathbf{v}_N) \tag{A5.1}$$

$$\frac{d\mathbf{r}_i}{dt}=\mathbf{v}_i \tag{A5.2}$$

where $\mathbf{F}_i(t,\mathbf{r}_1,...\mathbf{r}_N,\mathbf{v}_1,..., \mathbf{v}_N)$ is the force which acts on $i$ th particle.

We introduce the $3N$-dimensional configuration space $q=(\mathbf{r}_1,..., \mathbf{r}_N)$. The state of the system under consideration at each moment of time can be represented by a point in this configuration space. In the process of system evolution, this point moves, describing a certain line - the configuration trajectory. The $3N$-dimensional vector $V=(\mathbf{v}_1,..., \mathbf{v}_N)$ is the velocity of the system in the configuration space.

Let us consider a set of identical systems that differ in the initial conditions, but satisfy the requirement of single-valuedness: all systems that are at the same point of the configuration space at the same time must have the same velocity $V$. The considered set of the systems in the configuration space $q$ is the Hamilton-Jacobi ensemble. The Hamilton-Jacobi ensemble can be considered as a "gas" of non-interacting points which moves in the configuration space.

To describe such a Hamilton-Jacobi ensemble, as for ordinary gas, one can use either Lagrangian or Eulerian approach [16]. In the Lagrange approach, the motion of each individual Hamilton-Jacobi ensemble system in the configuration space is considered. This approach is connected with the material time derivative $\frac{d}{dt}$.

The equations of motion of the Hamilton-Jacobi ensemble in the Lagrangian approach coincide with Newton's equations (A5.1) and (A5.2). In the Eulerian approach, the Hamilton-Jacobi ensemble is viewed as a continuous medium which moves in a configuration space and has a certain configuration density $\rho(t,q)$ (the number of systems per unit volume of the configuration space) and a certain configuration velocity $V(t,q) = (\mathbf{v}_1,..., \mathbf{v}_N)$ at each point of the configuration space. In Eulerian approach, a local time derivative $\frac{\partial}{\partial t}$ is used that describes the change in the parameters of the Hamilton-Jacobi ensemble at a fixed point in the configuration space.

The connection between the material and local derivatives with respect to time is obtained by generalizing the relation (3):

$$\frac{d}{dt} = \frac{\partial}{\partial t} + (V\nabla)$$

or in expanded form

$$\frac{d}{dt} = \frac{\partial}{\partial t} + \sum_{i=1}^{N} (\mathbf{v}_i \nabla_i) \tag{A5.3}$$

where $\nabla_i = \frac{\partial}{\partial \mathbf{r}_i}$.

Substituting (A5.3) into the equation of motion (A5.1), one obtains the equation of motion of the Hamilton-Jacobi ensemble in the configuration space:

$$m_i \left( \frac{\partial \mathbf{v}_i}{\partial t} + \sum_{k=1}^{N} (\mathbf{v}_k \nabla_k) \mathbf{v}_i \right) = \mathbf{F}_i(t,q,V), \quad i = 1,...,N \tag{A5.4}$$

The density of the Hamilton Jacobi ensemble satisfies the continuity equation in the configuration space, which formally has the form

$$\frac{\partial \rho}{\partial t} + \mathrm{div}\,\rho V = 0$$

or in expanded form

$$\frac{\partial \rho}{\partial t} + \sum_{i=1}^{N} \nabla_i (\rho \mathbf{v}_i) = 0 \tag{A5.5}$$

Equations (A5.4) and (A5.5) generalize the Hamilton-Jacobi theory to arbitrary, including non-Hamiltonian, systems consisting of interacting particles.

Let us consider a system of particles interacting through the potential forces

$$\mathbf{F}_i(t,q) = -\nabla_i U(t,q) \tag{A5.6}$$

where $U = U(t,q)$ is the potential energy of the system of interacting particles.

In this case, the equation (A5.4) takes the form

$$m_i\left(\frac{\partial \mathbf{v}_i}{\partial t} + \sum_{k=1}^{N}(\mathbf{v}_k\nabla_k)\mathbf{v}_i\right) = -\nabla_i U \tag{A5.7}$$

It is easy to verify that equation (A5.7) has solutions corresponding to the potential flow of the Hamilton-Jacobi ensemble:

$$\mathbf{v}_i(t,q) = \frac{1}{m_i}\nabla_i S(t,q), \; i = 1,..., N \tag{A5.8}$$

herewith the function $S(t,q)$ satisfies the Hamilton-Jacobi equation for the system of interacting particles [1,2]

$$\frac{\partial S}{\partial t} + \sum_{k=1}^{N}\frac{1}{2m_k}|\nabla_k S|^2 + U = 0 \tag{A5.9}$$

The continuity equation in this case is obtained by substituting the relations (A5.8) into equation (A5.5).

## Acknowledgments

Funding was provided by the Tomsk State University competitiveness improvement program.

**Conflict of Interest:** The author declares that he has no conflict of interest.